\documentclass[mypaper,7pt,twoside]{CoAst}
\usepackage{epsf,graphicx,fancyhdr}
\renewcommand{\chaptermark}[1]{\markboth{#1}{}}

\newcommand{\kopf}{\small\itshape Comm. in Asteroseismology \\ Contribution to the Proceedings of the Wroclaw HELAS Workshop, 2008}

\newcommand{\Authors}[1]{\begin{center}\normalsize\bf\sf #1 \end{center}}

\renewcommand{\author}[1]{\begin{center}\normalsize\bf\sf #1 \end{center}}
\newcommand{\Address}[1]{\begin{center}\small\sf #1 \end{center}}

\newcommand{\Session}[1]{{\vspace{3mm}\small \noindent  \hspace*{3mm} Session: } #1 \normalsize}

\newcommand{\Objects}[1]{{\vspace{0mm}\small \noindent  \hspace*{3mm} Individual Objects: } \small #1 \normalsize}

	\newcommand{\threeA}{\small STARS - convection \newline}

\renewenvironment{abstract}{\section*{Abstract}\normalsize\sf}{}
\newcommand{\References}[1]{\begin{flushleft}{\large References\\}\vspace*{2mm}\small #1 \end{flushleft}}

\newcommand{\chapterCoAst}[2]{\chapter[\sf\normalsize #1\\ \footnotesize \hspace*{5mm}by #2 \sf\normalsize][]{#1\\}\rhead[\fancyplain{}{\sf\footnotesize \center{#1}}]{\fancyplain{}{\sffamily\thepage}}\lhead[\fancyplain{\kopf}{\sffamily\thepage}]{\fancyplain{\kopf}{\sf\footnotesize \center{#2}}}}

\newcommand{\figureCoAst}[5]{\begin{figure}[#4]
\centering
\includegraphics*[#5]{#1}
\caption{#2}
\label{#3}
\end{figure}}

\renewcommand{\chaptername}{}

\newcommand{\acknowledgments}[1]{\vspace*{5mm}\noindent  \textbf{Acknowledgments.} #1}

\def\rfr{\smallskip\par\noindent
        \hangindent=7truemm
        \hangafter=1}

\begin{document}
\sf

\chapterCoAst{The role of negative buoyancy in convective Cepheid models. Double-mode pulsations revisited}
{R.\,Smolec} 
\Authors{R.\,Smolec$^{1}$} 
\Address{
$^1$ Copernicus Astronomical Center, \\ Bartycka 18, 00-716 Warszawa, Poland
}

\noindent
\begin{abstract}
The longstanding problem of modeling double-mode behaviour of classical pulsators was solved with the incorporation of turbulent convection into pulsation hydrocodes. However, the reasons for the computed double-mode behaviour were never clearly identified. In our recent papers (Smolec \& Moskalik 2008a,b) we showed that the double-mode behaviour results from the neglect of negative buoyancy effects in some of the hydrocodes. If these effects are taken into account, no stable non-resonant double-mode behaviour can be found. In these proceedings we focus our attention on the role of negative buoyancy effects in classical Cepheid models. 
\end{abstract}

\Session{ \threeA }
\Objects{double-mode Cepheids} 

\section*{Introduction}

Since the early days of nonlinear pulsation computations, modeling double-mode (DM) phenomenon was one of the major objectives. However, the search for non-resonant double-mode behaviour with radiative hydrocodes failed (see however, Kov\'acs \& Buchler 1993). Typical modal selection observed, was, first overtone (1O) pulsation at the hot side of the instability strip (IS), fundamental mode (F) pulsation at the red side, and either-or domain (F/1O) in between. Incorporation of turbulent convection into pulsation hydrocodes led to stable and robust double-mode pulsation (Koll\'ath {\it et al.} 1998, Feuchtinger 1998). Most of the so far published double-mode models were computed with the use of the Florida-Budapest hydrocode ({\it eg.} Koll\'ath {\it et al.} 2002). This hydrocode adopts time-dependent convection model based on the Kuhfu\ss{} (1986) work. Kuhfu\ss{} model was also adopted in our pulsation hydrocodes (Smolec \& Moskalik 2008a), however, we could not find a double-mode behaviour, despite extensive search for it (Smolec \& Moskalik 2008b). We trace the difference in the computed modal selection to different treatment of negative buoyancy effects in both codes. In the Florida-Budapest code negative buoyancy is neglected, while it is present in our hydrocodes. Comparison of both treatments (performed with our hydrocodes) allow to understand the reasons for double-mode behaviour computed with hydrocodes that neglect negative buoyancy.

\section*{Turbulent convection model}

In Kuhfu\ss{} model of turbulent convection, equations of motion and energy conservation are supplemented with additional, single equation for generation of turbulent energy, $e_t$. Details of the model may be found {\it e.g.} in Smolec \& Moskalik 2008a. Here we focus our attention on turbulent energy equation and its crucial terms (turbulent pressure and flux of turbulent kinetic energy neglected):
\begin{equation}\frac{de_t}{dt}=S-D+E.V.\end{equation}  
In the above equation, $S$ is turbulent source function (or driving function), responsible for driving of turbulent eddies through the buoyant forces. Source function is proportional to superadiabatic gradient, $Y=\nabla-\nabla_{\mathrm{a}}$, and to the speed of convective elements, $\sim e_t^{1/2}$. The $D$-term describes the decay of turbulent eddies, through the turbulent cascade, $D\sim e_t^{3/2}$. $E.V.$ describes energy transfer through eddy-viscous forces. It describes the interplay between turbulent motion and mean gas motion, being proportional to the speed of convective elements and to the spatial derivative of the scaled mean velocity field, $U/R$. It always damps the pulsations and contributes to the driving of turbulent energy\footnote{Exact form of this term differ in different pulsation hydrocodes, see Smolec \& Moskalik 2008a, however results presented here are insensitive to these differences.}.

The crucial difference between our and Florida-Budapest formulation is the treatment of turbulent source function in convectively stable ($Y<0$) regions of the model. In our hydrocode we allow for negative values of source function, just as in the original Kuhfu\ss{} model ($S\sim Y$). In Florida-Budapest approach source term is restricted to positive values only ($S\sim Y_+$), which is equivalent to the neglect of negative buoyancy. Following convention introduced in Smolec \& Moskalik (2008a) we will denote the convective recipe and models ignoring negative buoyancy as PP models (Florida-Budapest approach), while convective recipe and models including negative buoyancy will be denoted by NN (our default formulation).

\section*{Consequences of neglecting negative buoyancy}

Using our pulsation hydrocodes we have performed a detailed comparison of models, differing only in the treatment of the source function (NN {\it vs.} PP models). Crucial differences are observed for single-mode limit cycle (full amplitude, monoperiodic oscillation) models. Amplitude of the models neglecting negative buoyancy (PP) is much lower than the amplitude of the models including negative buoyancy effects (NN). Lowering of amplitude in case of the PP models is connected with the eddy-viscous damping of pulsations in the deep convectively stable regions of the model. This damping is clearly visible in the nonlinear work integrals presented in Fig.~\ref{work}. For PP model a significant eddy-viscous damping below zone 70 (marked with arrows in the Figure), not present in NN model is visible. These internal zones are convectively stable ($Y<0$), however, in PP model significant turbulent energies are present in these zones, and hence eddy-viscous damping is possible. How these turbulent energies are built up? The bottom boundary of envelope convection zone, connected with hydrogen-helium ionization, is located roughly at zone 70 for both PP and NN models. Below this boundary, turbulent motions are effectively braked in NN model due to negative value of turbulent source function (eq.~1). Negative buoyancy slows down the turbulent motions very effectively and eddy-viscous damping is not possible in the inner parts of the model, due to negligible turbulent energies. Situation is different in case of PP models, in which turbulent source function is set equal to zero in convectively stable layers. Therefore, in these layers, turbulent energies are set by the balance between turbulent dissipation term ($D$ term in eq.~1), which damps the turbulent motions, and eddy-viscous term, which drives the turbulent energies. Due to neglect of negative buoyancy, turbulent motions cannot be braked effectively.  To the contrary, they are driven at the cost of pulsation, through the eddy-viscous term. Below the envelope convection zone turbulent energies are as high as $10^9-10^{10}$ erg/g -- only three orders of magnitude smaller than in the center of the convection zone. Such high turbulent energies extend to more than six local pressure scale heights below the envelope convection zone, leading to significant eddy-viscous damping, visible in the left panel of Fig.~\ref{work}. Deeper in the envelope, turbulent energies slowly decay, reflecting the vanishing amplitude of the pulsations.

\figureCoAst{smolec1.eps}{Nonlinear work integrals plotted versus the zone number. Surface at right.}{work}{t}{clip,angle=0,width=115mm}

To check the effect of neglecting negative buoyancy on modal selection, we have computed several sequences of nonlinear Cepheid models using both PP and NN convective recipes. Details of mode selection analysis may be found in Smolec \& Moskalik (2008b). Each model was initialized (kicked) with several different initial conditions (mixtures of F and 1O linear velocity eigenvectors) and time evolution of the fundamental mode and first overtone amplitudes, $A_0$ and $A_1$, was followed. Exemplary results are shown in Fig.~\ref{traj}. Computed trajectories are plotted with solid lines for PP model and dotted lines for NN model. Hydrodynamic computations coupled with amplitude equation analysis allow to find all stable pulsation states to which trajectories converge (attractors, solid squares in Fig.~\ref{traj}), and all unstable solutions, that repel the trajectories (open squares in Fig.~\ref{traj}). For single-mode solutions stability coefficients are computed. These are $\gamma_{1,0}$, which describes the stability of the fundamental mode limit cycle with respect to first overtone perturbation (switching rate toward 1O) and $\gamma_{0,1}$, which measures the stability of the first overtone limit cycle. Negative value of $\gamma$ means that the respective limit cycle is stable. If both coefficients, $\gamma_{1,0}$ and $\gamma_{0,1}$ are simultaneously positive, both limit cycles are unstable, and double-mode pulsation is unavoidable. The run of stability coefficients across the instability strip, for sequence of Cepheid models, computed with both PP and NN convective recipes, is presented in Fig.~\ref{stab}. Arrow in this Figure marks the location of model, for which hydrodynamic integrations are presented in Fig.~\ref{traj}.

\figureCoAst{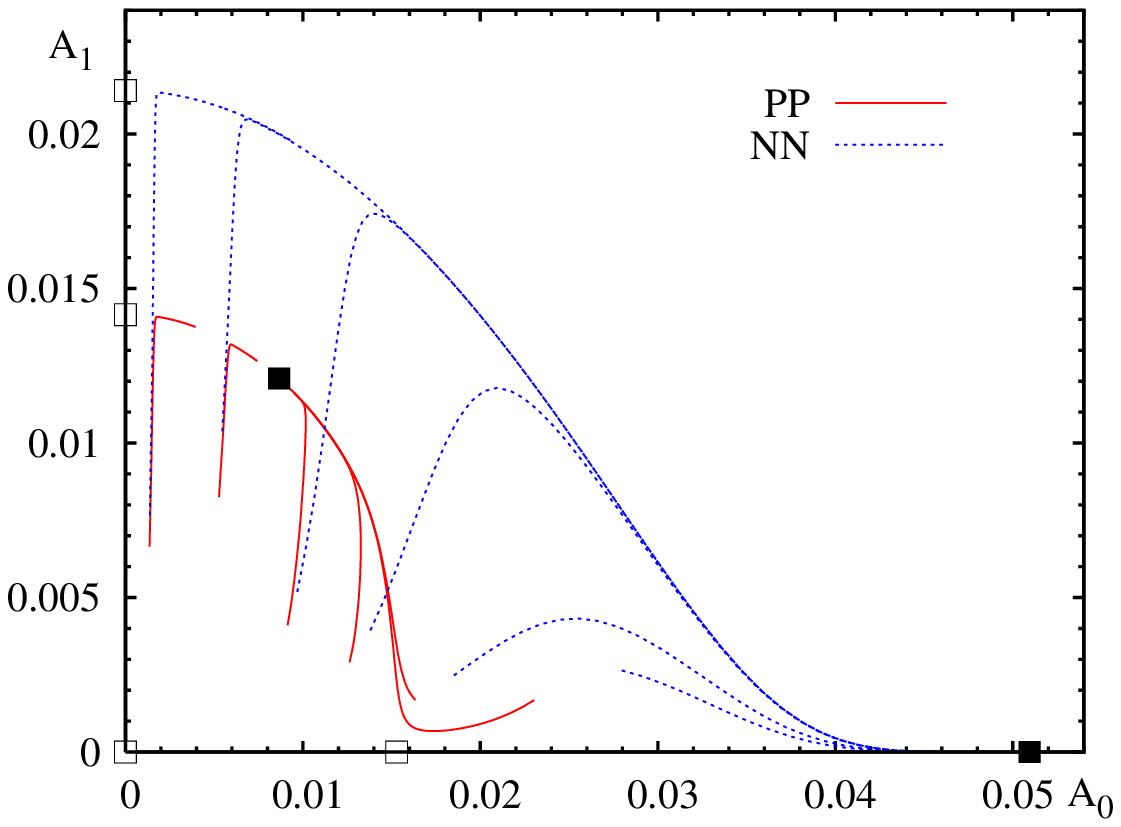}{Hydrodynamic integrations for a particular Cepheid model. Solid lines - model computed with PP convection, dotted lines model computed with NN convection.}{traj}{h}{clip,angle=0,width=80mm}
\figureCoAst{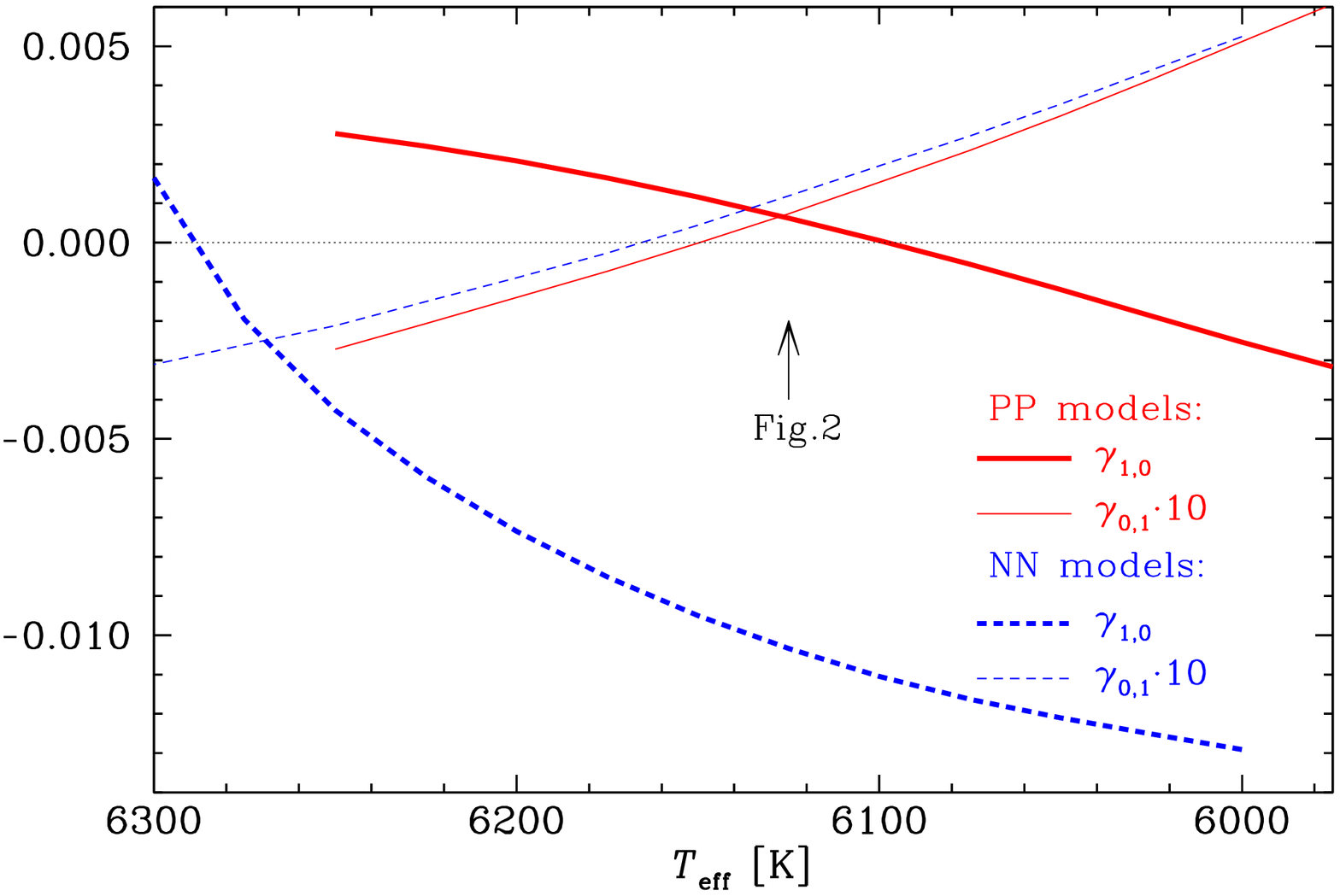}{The run of stability coefficients along a sequence of Cepheid models of constant mass/luminosity. Solid lines for PP models, dashed lines for NN models.}{stab}{h}{clip,angle=0,width=80mm}

Amplitude of the given mode is a main factor affecting its stability. Higher the amplitude of the mode, more able it is to saturate the pulsation instability alone, and hence, more stable its limit cycle is. For NN models amplitude of the fundamental mode is much higher than the amplitude of the first overtone (see Fig.~\ref{traj}) across significant part of the instability strip. Therefore, stability coefficient of the fundamental mode limit cycle, $\gamma_{1,0}$, becomes negative very close to the blue edge of the IS at temperature $\approx 6290$K (Fig.~\ref{stab}). Double-mode state is not possible, as first overtone becomes unstable ($\gamma_{0,1}>0$) for much lower temperature ($T\approx 6165$K). In a temperature range in which both limit cycles are stable, either-or domain is observed, in which pulsation in either limit cycle is possible.

In case of PP models, amplitudes of both modes are reduced as compared to NN models, but not by the same amount. As is well visible in Fig.~\ref{traj}, amplitude of the fundamental mode is reduced much more than the amplitude of the first overtone. This effect is explained by higher amplitude of the fundamental mode (in comparison to 1O) in convectively stable layers of PP model, in which significant turbulent energies are observed. As a result, eddy-viscous damping is stronger for the fundamental mode. Differential reduction of mode amplitudes is crucial in bringing up double-mode behaviour in PP model sequence. At the hot side of the IS, amplitude of the fundamental mode is lower, or comparable to the amplitude of the first overtone. Hence, fundamental mode limit cycle is unstable ($\gamma_{1,0}>0$) in a wide temperature range. It becomes stable at much lower temperature ($T\approx 6100$K) in comparison to NN sequence. At this temperature first overtone is already unstable. Consequently, in relatively wide temperature range ($\sim 50$K in Fig.~\ref{stab}) both limit cycles are unstable and double-mode state emerges.

\section*{Summary}

Neglect of negative buoyancy has serious consequences for the computed Cepheid models. It leads to high turbulent energies in convectively stable layers, and consequently to strong eddy-viscous damping. This damping acts differentially on pulsation modes, which promotes the double-mode behaviour. If buoyant forces are taken into account, as they should be, no stable non-resonant double-mode behaviour can be found (Smolec \& Moskalik 2008b). Therefore, the problem of modeling F/1O double-mode behaviour in classical Cepheids remains open.

\acknowledgments{I'm very grateful for many fruitful discussions with Pawel Moskalik and Wojciech Dziembowski. Pawel Moskalik is acknowledged for reading the manuscript. I am grateful to the referee, Geza Kov\'acs, for many comments on the presented results. The author is grateful to the EC for the establishment of the European Helio- and Asteroseismology Network HELAS, as well as for financial support, which made the participation of the author at this workshop possible. This work has been supported by the Polish MNiSW Grant No. 1 P03D 011 30.}

\References{
\rfr Feuchtinger, M. 1998, Astron. Astrophys., 337, 29
\rfr Koll\'ath, Z., Beaulieu, J.P., Buchler, J.R., et al. 1998, Astrophys. J., 502, L55
\rfr Koll\'ath, Z., Buchler, J.R., Szab\'o, R., et al. 2002, Astron. Astrophys., 385, 932

\rfr Kov\'acs, G. \& Buchler, J.R. 1993, Astrophys. J., 404, 765
\rfr Kuhfu\ss{}, R. 1986, Astron. Astrophys., 160, 116
\rfr Smolec, R., \& Moskalik, P. 2008a, Acta Astron., 58, 193
\rfr Smolec, R., \& Moskalik, P. 2008b, Acta Astron., 58, 233
}

\end{document}